\documentclass{aastex631}
\usepackage{color}
\usepackage{hyperref}
\usepackage{epstopdf}
\usepackage{graphicx}
\usepackage[FIGTOPCAP]{subfigure}
\usepackage{amsmath}
\usepackage{ulem}

\newcommand{\tdiff}{t_{\rm diff}}
\newcommand{\Erot}{E_{\rm rot}}
\newcommand{\Eem}{E_{\rm EM}}
\newcommand{\Edotrot}{\dot{E}_{\rm rot}}

\newcommand{\Lem}{L_{\rm EM}}
\newcommand{\Lemo}{L_{\rm EM,0}}
\newcommand{\Lgw}{L_{\rm GW}}
\newcommand{\Lrad}{L_{\rm rad}}

\newcommand{\Mej}{M_{\rm ej}}

\shorttitle{GW radiation from magnetar-driven supernovae}
\shortauthors{Xie et al. }

\begin{document}
\title{Gravitational wave radiation from the magnetar-driven supernovae}

\correspondingauthor{Lang Xie}
\email{xielang@nao.cas.cn}

\author{Lang Xie}
\affiliation{National Astronomical Observatories, Chinese Academy of Sciences, Beĳing 100101, China}
\affiliation{School of Astronomy, University of Chinese Academy of Sciences, Beĳing 100049, China}

\author{Hong-Yu Gong}
\affiliation{Key Laboratory of Dark Matter and Space Astronomy, Purple Mountain Observatory, Chinese Academy of Sciences, Nanjing 210034, China}
\affiliation{School of Astronomy and Space Science, University of Science and Technology of China, Hefei, Anhui 230026, China}
\author{Long LI}
\affiliation{Department of Astronomy, University of Science and Technology of China, Hefei 230026, China}
\author{Da-Ming Wei}
\affiliation{Key Laboratory of Dark Matter and Space Astronomy, Purple Mountain Observatory, Chinese Academy of Sciences, Nanjing 210034, China}
\affiliation{School of Astronomy and Space Science, University of Science and Technology of China, Hefei, Anhui 230026, China}
\author{J. L. Han}
\affiliation{National Astronomical Observatories, Chinese Academy of Sciences, Beĳing 100101, China}
\affiliation{School of Astronomy, University of Chinese Academy of Sciences, Beĳing 100049, China}

\begin{abstract}
Rapidly spinning magnetars are potential candidates for the energy source of supernovae (SNe) and gamma-ray bursts and the most promising sources for continuous gravitational waves (GWs) detected by ground-based GW detectors. Continuous GWs can be radiated from magnetars due to magnetic-induced deformation or fluid oscillations, compatible with magnetic dipole (MD) radiation for spin-down energy. In this paper, we investigate the diverse light curves of magnetar-driven SNe in the scenario that the spin-down is dominated by GW radiation and/or MD radiation. By simulating the light curves of SNe and employing the Markov chain Monte Carlo method, we constrain the parameters of the magnetars and SN explosions and show that the signature of GW radiation may be indicated by the bolometric luminosity curves of SNe Ic-BL 2007ru and 2009bb. We find that the ellipticity of magnetars in the order of $10^{-3}$ can be induced by the magnetic field of $\sim 10^{16} \mbox{ G}$. 
If such continuous GWs associated with SNe can be detected in the future by the Advanced LIGO and Virgo detectors, this would be a smoking gun for a magnetar engine powering SNe.
\end{abstract}

\keywords{gravitational waves --- supernovae: general --- stars: magnetars}

\section{ Introduction }
A long-lived magnetar can be formed by the core collapse of a massive star, which can produce a transient event, such as a gamma-ray burst (GRB) and a supernova (SN). The rotational energy of the magnetar is a reservoir for the afterglow emission of SN and GRB with diverse evolutionary features, such as short GRB having extended emission or GRB with X-ray plateau \citep{Usov1992, Dai1998a, Dai1998b, Zhang2001, Metzger2011, Gompertz2013, Rowlinson2013, Metzger2014, L2014, Sarin2020, Li2021, xie2022, Xie2022b}. The high kinetic energy or high luminosity SNe are generally interpreted as an extra injection of spin-down energy from the magnetar engine  \citep{Kasen2010, Woosley2010, Inserra2013, Metzger2015, Kashiyama2016, Kasen2016, Wang2016b, Moriya2017, Nicholl2017, Omand2024}. 

Many observations show that some broad-lined Type Ic (Ic-BL) SNe and superluminous SNe (SLSNe) have peak luminosities and radiant energies tens of times higher than those of typical SNe \citep{Gal2012, Nicholl2021}. Various models have been applied in the literature to interpret the striking kinetic energy and luminosity. One popular model involves the energy injection from a rapidly spinning magnetar \citep{Woosley2010, Kasen2010}. The basic picture is that winds from a magnetar driven by magnetic dipole radiation interact with the SNe ejecta and deposit their spin energy into the SN ejecta. Such a magnetar-driven model has successfully explained the light curves of SNe Ic-BL \citep{Greiner2015, Cano2016, Wang2016b} and SLSNe (i.e, \citep{Dessart2012, Chatzopoulos2013, Inserra2013, Nicholl2013, Howell2013, Nicholl2017, Yu2017, Liu2017, Villar2018, Wang2019}), as well as the nebular spectral observations of hydrogen-poor SLSNe \citep{Jerkstrand2017A, Quimby2018, Nicholl2019}. Other models for powering the energetic SNe involve the radioactive decay of massive amounts of $^{56}$Ni \citep{Barkat1967, Heger2002, Gal2009}, the jets launched by the center engine for SNe Ic-BL \citep{Lazzati2012, Gilkis2016, Barnes2018, Eisenberg2022, Corsi2023}, the interaction of SN ejecta with the circumstellar medium \citep{Blinnikov2010, Chevalier2011, Ginzburg2012, Moriya2012}, and an accreting black hole (i.e., \citep{MacFadyen2001, Dexter2013, Li2020}).

In the magnetar-driven SNe scenario, a magnetar with a strong magnetic field of $B\sim 10^{15}\mbox{ G}$, a rapid rotation of initial spin period $P_0\sim 1\mbox{ ms}$, and high temperatures is born after the core collapse of massive star \citep{Duncan1992}. The rapid rotation of the magnetar can induce a large deformation due to the magnetic pressure or fluid oscillations \citep{Owen1998, Lai1995, Bonazzola1996, Andersson1998, Lindblom1998, Cutler2002, Stella2005, Haskell2008, Alford2015, Lasky2017}, implying that the magnetar could emit the remarkable GWs. The GW quadrupole radiation and magnetic dipole (MD) radiation torque could act together for magnetar spin-down \citep{Shapiro1983, Zhang2001}. The GW energy loss would therefore affect the spin evolution and also electromagnetic outflows of the magnetar. In other words, if a deformed magnetar powers an SN, the GW-energy loss could result in less rotational energy supplied to the SNe, and then lead to a change in the shape of the light curve, indicated by a lower peak luminosity and a slower decay rate. A magnetar-driven system should produce a unique luminosity evolution, different from only MD radiation losses.

Non-axisymmetrically deformed magnetars are potential continuous GW sources for the Advanced Laser Interferometer gravitational wave Observatory (aLIGO) detector \citep{Aasi2015} and the Advanced Virgo detector \citep{Acernese2015}. The aLIGO has searched for continuous GWs from various types of isolated neutron stars (NSs), such as known pulsars \citep{abbott2019a}, post-burst magnetars \citep{abbott2019d, abbott2022}, accreting NSs and young NSs in SN remnants \citep{abbott2019b, abbott2019f}. However, limited by the sensitivity at the designed frequencies, no GWs have yet been detected from these newly formed compact objects. 

Detecting GWs from magnetar-driven SNe could help us constrain the structure and composition of NSs and provide theoretical traction for future ground-based GW telescopes. We explore in this paper possible indications of the GW radiation in SNe electromagnetic light curves. We develop a model for magnetar-driven SNe by considering the GW radiation and the effect on the bolometric luminosity evolution of SNe.  From the SN luminosity curves, we can constrain the characteristics of magnetars, i.e., the initial spin period $P_0$, the dipole magnetic field strength $B$, and the ellipticity of the magnetar $\epsilon$, as well as the parameters of SN explosions. Through the best-fitting for parameters of the magnetar obtained by the Markov chain Monte Carlo (MCMC) method, we derive the GW characteristic amplitude of the magnetar and then evaluate the GW detectability of future ground-based GW telescopes. This paper is organized as follows. We investigate the effect of the GW energy-less on the bolometric luminosity evolution of SNe under the assumption of the magnetar model in section 2. In section 3, we take some SNs as examples and fit their SN observation data with the magnetar model. In Section 4, we analyze the detectability of GWs from magnetars that power SNe. The discussion and conclusions are given in Section 5.

\section{A magnetar model with gravitational wave emission}

Some authors \citep{Inserra2013, Nicholl2017} have used magnetar models to explain SNe light curves and constrain the parameters of magnetars and the ejecta mass.  However, the effect of GW radiation is usually neglected. %The main objective of this paper is to search for the signature of GW radiation in the electromagnetic observations of SNe and to predict the detectability of their GWs. 

A rapidly spinning magnetar can be produced after the core collapse of a massive star. The rotational energy of the magnetar provides a huge energy reservoir for SNe, which greatly enriches the features of the SNe optical emission. However, not all of the rotational energy is used to heat the SN ejecta, part of which may be lost via the significant GW radiation because of non-axisymmetric deformation. Thus, the total rotational energy of the magnetar, $\Erot = I\Omega^2/2$, is released through both GW radiation and MD radiation, and the spin-down evolution can then be written as \citep{Shapiro1983, Zhang2001}
\begin{equation}
\Edotrot = I\Omega\dot{\Omega} =\Lem + \Lgw,
\label{eq:spin}
\end{equation}
where $I$ is the moment of inertia of a neutron star, $\Omega$ and $\dot{\Omega}$ are the spin velocity and its time derivative. The electromagnetic luminosity $\Lem$ generated by the magnetar spin-down is given by \cite{Spitkovsky}) 
\begin{equation}
\Lem = -\frac{B^2R^{6}\Omega^{4}\sin^2\theta}{6c^{3}},
\label{eq:Lem}
\end{equation} 
Here $B$ is the magnetic dipole field strength on the neutron star surface, $\theta$ is the angle between the magnetic axis and the spin axis, and $R$ and $c$ are the radius of the NS and the speed of light, respectively. In this paper, we adopt the typical parameters for a neutron star with a mass of $M=1.4\, M_{\sun}$, a radius of $R=10$~km, and the moment of $I=1\times 10^{45} {\rm g cm^2}$, and also assume that the neutron star has evolved to an orthogonal rotator ($\theta =\pi/2$).

Various causes for the NS deformation have been proposed in the previous literature. First, super-strong magnetic fields exist inside the NS due to differential rotation and magnetic dissipation instability. The anisotropic pressure generated by strong magnetic fields then deforms the NS into an ellipsoid \citep{Bonazzola1996, Cutler2002, Stella2005}. It has been argued that magnetic-induced deformation plays an important role in the early evolution of magnetars \citep{Corsi2009, Dall'Osso2015, Lasky2017, Araujo2016}.  Second, the NS accretion can also lead to the deformation \citep{Haskell2015, Zhong2019, Sur2021}. The accreted material flows towards the poles of the NS, compresses and obscures the local magnetic field, and thus forms ``mountains'' on the surface of the NSs. Third, the rapid spin of young NSs undergoes unstable oscillatory modes that couple to the gravitational field \citep{Andersson1998, Lindblom1998, Alford2015}. The spinning magnetars with non-axisymmetric deformations possess quadrupole moments, which would emit significant GW as being \citep{Usov1992, Zhang2001})
\begin{equation}
\Lgw = -\frac{32}{5}\frac{GI^2}{c^5}\epsilon^2\Omega^6.
\label{eq:Edotgwe}
\end{equation}
Here $G$ is the gravitational constant, and $\epsilon$ is the ellipticity of the star. 
Considering the Eq.\ref{eq:spin}, one can find the spin velocity $\Omega$ evolving:
\begin{equation}
\frac{d\Omega}{dt} = -\alpha\Omega^3-\beta\Omega^5, \label{eq:omegadot}
\end{equation}
where $\alpha\equiv B^2R^6/6c^3I$, $\beta\equiv 32GI\varepsilon^2/5c^5$. The evolution of spin velocity $\Omega$ depends on the combination of MD radiation and GW radiation. Here are three possible kinds of spin-down evolution of magnetars: 
(1) If the magnetar spins down only due to MD radiation and the contribution of GW radiation is negligible, the electromagnetic luminosity evolves as $\Lem= \Lemo  (1+t/\tau_{\rm em})^{-2}$, where $\tau_{\rm em}=\frac{1}{2\alpha\Omega_0^2}$ is the timescale of MD radiation. The electromagnetic luminosity shows a plateau at $t<\tau_{\rm em}$, then decays as $\Lem\propto t^{-2}$ for $t>\tau_{\rm em}$. 
(2) If the GW radiation dominates the spin-down, the electromagnetic luminosity can be expressed as $\Lem= \Lemo(1+t/\tau_{\rm gw})^{-1}$, where $\tau_{\rm gw}= \frac{1}{2\beta\Omega_0^4}$ is the timescale of GW radiation. The electromagnetic luminosity also exhibits a plateau before decaying as $\Lem\propto t^{-1}$ for $t>\tau_{\rm gw}$.  
(3) If both the GW radiation and MD radiation act together for the magnetar spin-down, the electromagnetic luminosity would show a change from $\Lem\propto t^{-1}$ to $\Lem\propto t^{-2}$ after the plateau phase. The radiation efficiency of GW and MD depends on the spin frequency (i.e., GW for $\Omega^{6}$ vs MD for $\Omega^{4}$), so the GW radiation should dominate the early evolution of the electromagnetic luminosity and the MD dominates the later evolution, causing the diversity of magnetar-driven SNe.

With the above basic ideas, we simulate how magnetars interact with SNe ejecta in different spin-down processes and how the different SNe light curves can be found. Our models are built based on the magnetar-driven SNe theory proposed by \citet{Kasen2010}(see also \cite {Kashiyama2016, Ho2016, Omand2024} and \citet{Kasen2016}).

\begin{figure}
 \centering
\includegraphics[width=0.6\textwidth]{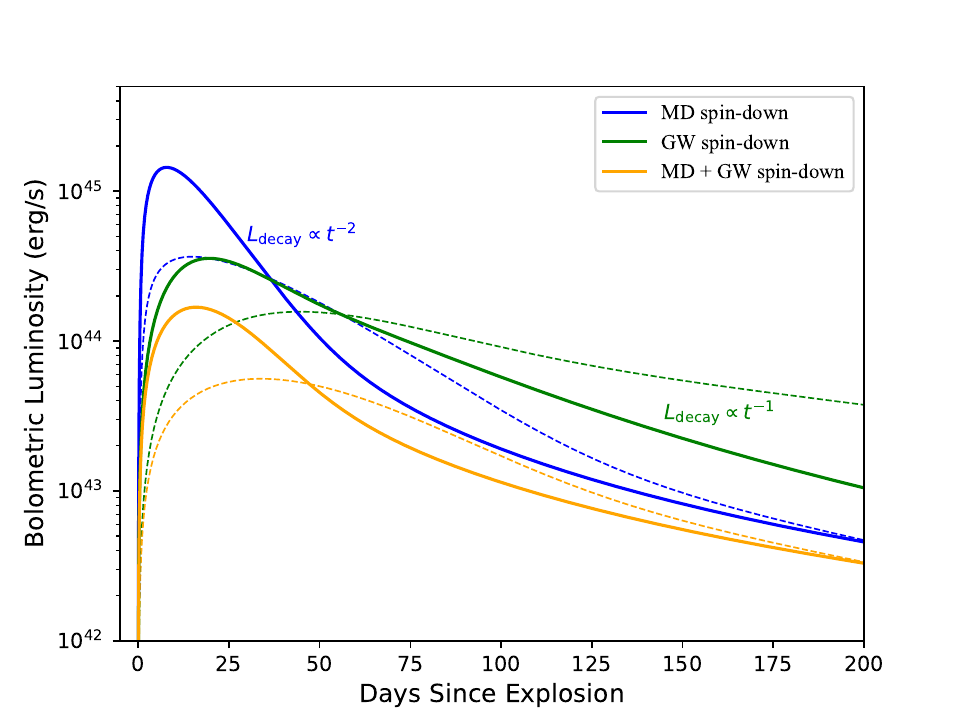}
\caption{The evolution of bolometric luminosity for a magnetar-driven supernovae in different spin-down modes, simulated with parameters of $B=1\times10^{14}\mbox{ G}$,  $P_0=1\mbox{ ms}$,  $\epsilon = 2\times 10^{-3}$,  $\kappa$ = 0.1 cm$^2$~g$^{-1}$, ${\kappa_\gamma}$ = 0.1 cm$^2$~g$^{-1}$, $E_{\rm sn}=10^{51}\mbox{ erg}$, $M_{\rm ej}=1\, M_{\sun}$ (solid lines) and $M_{\rm ej}=3\, M_{\sun}$ (dashed lines). The blue solid and dashed lines are plotted for the MD radiation spin-down; the green solid and dashed lines for the GW radiation spin-down; and the orange solid and dashed lines by both GW and MD radiations. }
\label{fig-bolo}
\end{figure}

A relativistic wind generated by the MD radiation from the magnetar goes outward and heats the SN ejecta. Approximating the ejecta at time $t$ in the form of a uniform sphere with a mass of $M_{\rm ej}$ and a radius of $R_{\rm ej}$, with a velocity of $v_{\rm sc}$ in the volume of $V \approx 4\pi R_{\rm ej}^{3}/3$ and the density of $\rho = M_{\rm ej}/V$, one can describe the evolution of the internal energy as being \citep{Kasen2016}
\begin{equation}
\frac{\partial E_{int}}{\partial t} =  \xi \Lem - \Lrad- P\frac{ \partial V}{\partial t}.
 \label{eq:energy}
\end{equation}
The first term on the right is the electromagnetic luminosity injected by the magnetar due to spin-down. Here $\xi$ is the efficiency of thermalization of electromagnetic energy into radiation. The second term is the thermal radiation luminosity of SNe. The third term is the energy loss due to the expansion of the ejecta, $P$ is the radiation pressure that is related to the internal energy by $P = E_{int}/3V$,  ${\partial V}/{\partial t} =  4\pi R_{\rm ej}^{2}v_{\rm sc} $, and then $-P({ \partial V}/{\partial t})=E_{\rm int}/t$.
The $\xi$ evolves with time and is related to the gamma-ray leakage of the ejecta. Hence, we denote $\xi$ as 
\begin{equation}
   \xi = 1 - e^{-At^{-2}},
   \label{eq:xi}
\end{equation}
where 
\begin{equation}
    A = \frac{3 \kappa_\gamma M_{\rm ej}}{4\pi v_{\rm sc}^2}
    \label{eq:leakage}
\end{equation}
is parameter of gamma-ray leakage \citep{Wang2015b}, $\kappa_\gamma$ represents the opacity of gamma-ray.
The radiated luminosity of SN is given by \citep{Kasen2010, Kotera2013}
\begin{eqnarray}
    L_{\rm rad}  =& \frac{E_{int} c}{\tau R_{\rm ej}} = \frac{E_{int} t}{t_{\rm diff}^2}, {\rm ( \tau < 1), }\label{eqn:lrad_p} \\=& \frac{E_{int} c}{R_{\rm ej}},   {\rm (\tau > 1)},\label{eq:lrad_l}
\end{eqnarray}

where the optical depth of the ejecta is $\tau = 3\kappa M_{\rm ej}/{4\pi R_{\rm ej}^{2}}$, and the radiative diffusion timescale is
\begin{eqnarray}
\tdiff &=& \left(\frac{3\kappa\Mej}{4\pi cv_{\rm sc}}\right)^{1/2}, 
\label{eq:tdiff}
\end{eqnarray}
here $\kappa$ is the opacity of ejecta.
The expansion velocity of the ejecta is given by \citep{Kasen2010}
\begin{eqnarray}
v_{\rm sc} & =& \left[\frac{2(\Eem(\Omega) +E_{\rm sn})} {\Mej}\right]^{1/2}, 
\label{eq:velo}
\end{eqnarray}
where $E_{sn}$ is the initial kinetic energy of ejecta, $\Eem$ is the energy of MD radiation integrated $\Lem$ over time.

\begin{figure}
 \centering
\includegraphics[width=0.49\textwidth]{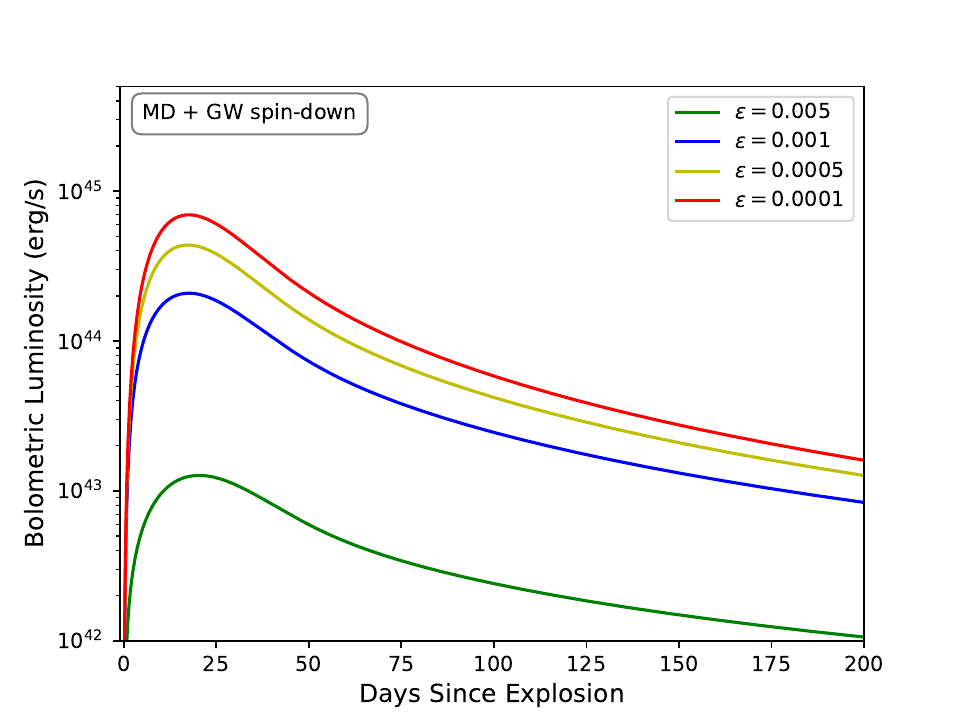}
\includegraphics[width=0.49\textwidth]{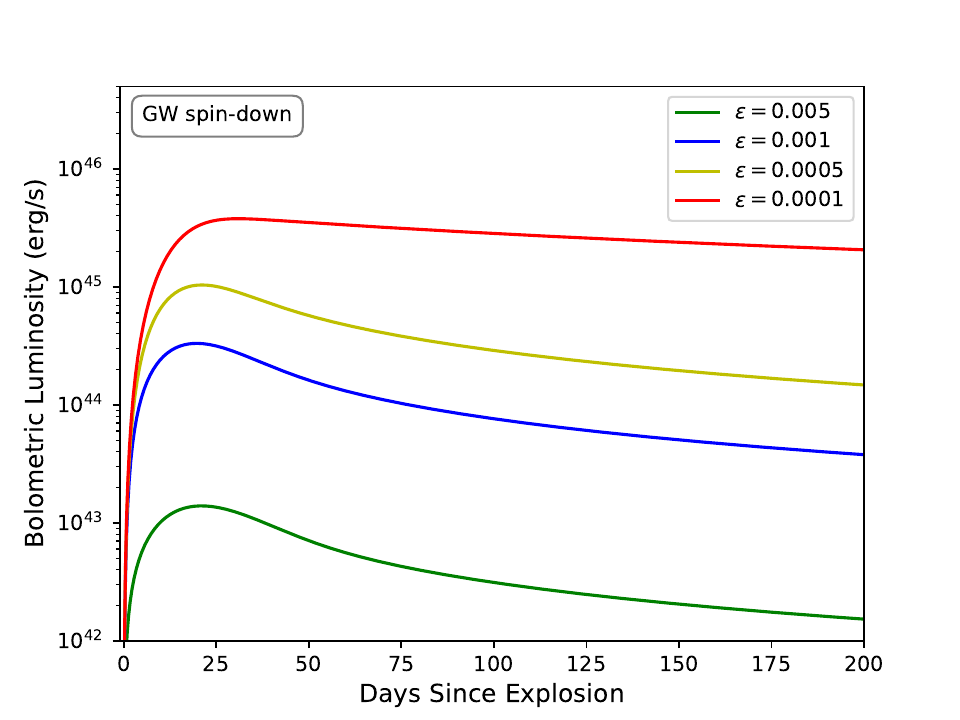}
\caption {The same as Figure~\ref{fig-bolo} but the evolution of supernova bolometric luminosity caused by different NS ellipticities for the spin-down by both the GW and MD radiations or by merely the GW radiation.}
\label{fig-ellip}
\end{figure}

\citet{Kasen2010} and \citet{Woosley2010} were the first to develop a simplified model of magnetar-driven SNe and showed that the magnetar models can reproduce the timescales and brightness of SNe within reasonable parameters. Modeling the light curves of magnetar-driven SNe has to choose the magnetar spin evolution channel and get the solution of the internal energy evolution and the velocity of ejecta. Fig \ref{fig-bolo} demonstrates the magnetar-driven SNe light curves for three different spin-down modes, the magnetar MD radiation spin-down (blue line), the GW radiation spin-down (green line), and both (orange line), calculated for the initial parameters such as the magnetic field strength of $B=1\times10^{14}\mbox{ G}$, the initial spin period of $P_0=1\mbox{ ms}$, the ellipticity of $\epsilon = 2\times 10^{-3}$, the opacity of $\kappa$ = 0.1 cm$^2$~g$^{-1}$, and the gamma-ray opacity of ${\kappa_\gamma}$ = 0.1 cm$^2$~g$^{-1}$, the initial explosion energy of $E_{\rm sn}=10^{51}\mbox{ erg}$, and the ejecta mass of $1\,M_{\sun}$ (solid lines) and $3\,M_{\sun}$ (dashed lines). 

In magnetar-driven SNe scenarios, the light curve evolution after its peak luminosity can be described as $L_{rad}\propto t^{-\alpha }$. The value of $\alpha$ depends on the declining index of the electromagnetic luminosity injected by the magnetar. Often $\alpha = 2$ is taken for a magnetar to spin down due to only the MD radiation, and $\alpha = 1$ for the GW radiation spin-down, or some values in between for the combined cases. 

The evolution of the bolometric luminosity of the SN differs due to the ellipticity
of a deformed NS, as shown in Fig \ref{fig-ellip}. A larger ellipticity would cause a lower peak luminosity of the SN. The greater the loss of the GW energy, the smaller the amount of electromagnetic energy provided to the SN for a smaller luminosity peak. In other words, the low-luminosity SNe is more likely to have a previously unknown GW radiation. 
When the MD radiation is involved for NSs with strong magnetic fields, only a large ellipticity can cause a significant decrease of the bolometric luminosity peak.  
The GW energy loss causes a dwarf peak of the SNe light curve and a slower decay rate of the SNe light curve tail. Note, however, that the observed $t^{-2}$ tail in some SN light curves cannot be directly considered as a signature of the presence of a magnetar engine, since the $^{56}$Co radioactive decay could exhibit a similar tail \citep{Inserra2013}.

\section{Fitting with supernova light curves}

We construct a magnetar model containing three spin-down modes and obtain the best-fitting parameters and the posterior parameter distributions by employing the emcee python package and the MCMC method \citep{Foreman2013}. The model contains the following seven free parameters: dipole magnetic field strength $B$, initial spin period $P_0$, NS ellipticity $\epsilon$, ejecta mass $M_{\rm ej}$, initial SN explosion energy $E_{\rm sn}$, opacity $\kappa$ and gamma-ray opacity $\kappa_{\gamma}$. The prior distribution of the parameters is set in the form of log-uniform or uniform distributions to ensure the parameters are in sufficiently large intervals, as seen in Table \ref{parameters}. The values of ${\rm \chi^{2}}/{\rm dof}$ are adjusted for the best-fitting model of the SN light curves. 

Three criteria are adopted for the choice of observation data. First, there must be sufficient observational data during the rise and decay of the light curve, and the observational data needs to extend beyond $100$ days. Second, the later phase of the light curve exhibits a slow decay tail with a decay rate between $-1$ to $-2$. This feature is consistent with what we expected from magnetar models, where enlarged decay rates for more rotational energy loss due to the GW radiation. Finally, SNe with the complex evolved behaviors in the decay phase are excluded from our sample. We here work on two cases, SN 2009bb and SN 2007ru.

\begin{table*}
%\centering
\tabletypesize{\scriptsize}
\caption{The free parameters, units, and prior distributions of the magnetar model and the best-fitting results of the SNe 2007ru and 2009bb. }
\label{parameters}
\hspace{-30pt}
\begin{tabular}{c c c c c c c c}
\hline\hline
                                & \colhead{Definition}                                 &  \colhead{Unit}      &    \colhead{Prior}   & \colhead{Values of SN 2007ru} & \colhead{Values of SN 2009bb}\\
\hline

{\bf Magnetar }\\
\hline
$B$                             & magnetic field strength           &  $10^{14}$ G         &    $[0.1, 50]$      & $12.55^{+0.32}_{-0.30}$ & $13.52^{+0.40}_{-0.37}$ \\
$P_0$                           & initial spin period                    &   ms                 &    $[1.0, 40]$       & $25.19^{+0.50}_{-0.51}$     & $26.86^{+0.45}_{-0.45}$ \\
$\epsilon$                     & ellipticity           &      $10^{-3}$                &    $[0.1, 20]$   & $4.79^{+3.58}_{-3.16}$  & $5.33^{+3.29}_{-3.58}$ \\
$M_{\rm ej}$                    & ejecta mass                                      &   M$_\odot$          &    $[0.1, 30]$      & $1.51^{+0.74}_{-0.48}$ & $1.08^{+0.96}_{-0.53}$ \\
$E_{sn}$                             &initial kinetic energy of ejecta                               &   $10^{50}$ cm s$^{-1}$ &    $[0.1, 100.0]$     & $29.08^{+14.21}_{-16.62}$ & $28.79^{+14.92}_{-16.80}$ \\
$\kappa$  &  optical opacity  &   cm$^2$g$^{-1}$     &    $[0.01, 0.2] $     & $0.03^{+0.03}_{-0.01}$ & $0.04^{+0.05}_{-0.02}$ \\
$\log \kappa_{\rm \gamma}$ & gamma-ray opacity                &   cm$^2$g$^{-1}$     &    $[-2, 2]$        &  $1.13^{+0.27}_{-0.48}$ & $0.72^{+0.53}_{-0.67}$ \\
$\chi^{\rm 2}$/dof              &                                                      &                      &                      & $32/23$ & $101/43$  \\
\hline\hline

\noalign{\smallskip}
\end{tabular}
\end{table*}

\subsection{SN 2009bb}

\begin{figure}
 \centering
\includegraphics[width=0.49\textwidth]{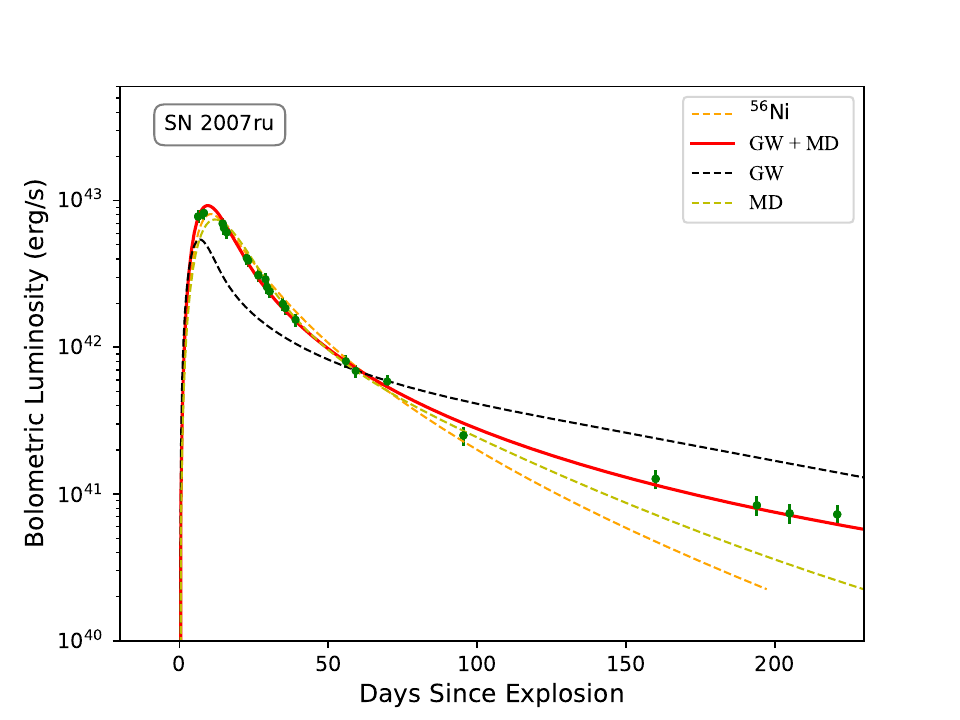}
\includegraphics[width=0.49\textwidth]{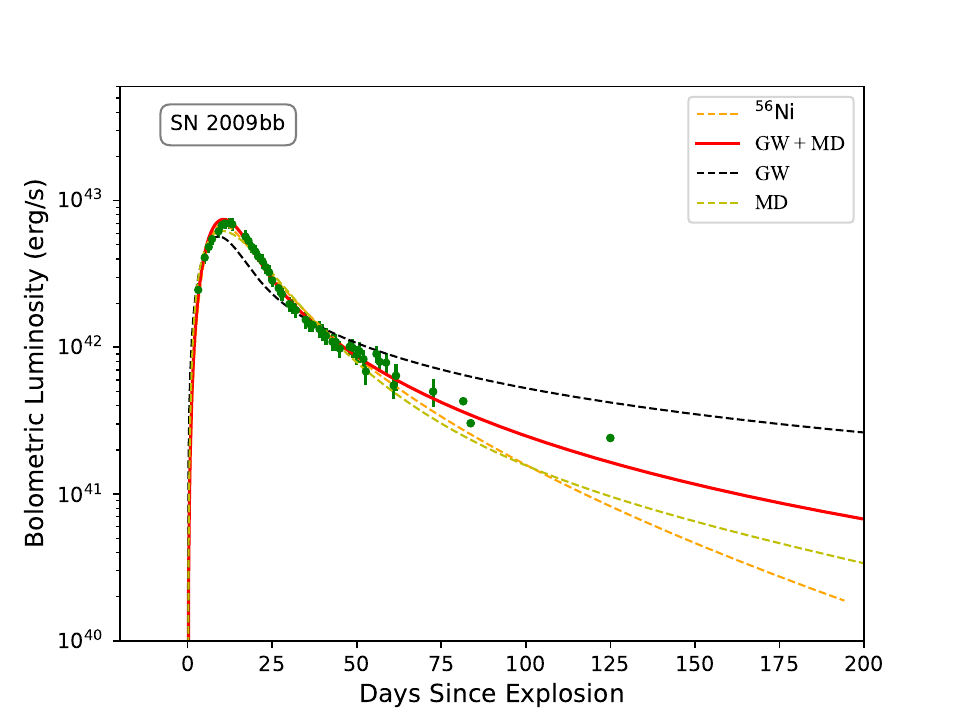}
\caption {Bolometric luminosity curves of SNe 2007ru and 2009bb together with model fits. Red solid lines represent the fitting results of the magnetar model with both GW and MD acting together, yellow dashed lines and black dashed lines indicate the fitting results of the magnetar models for the MD and GW spin-downs, respectively. The results of the $^{56}$Ni model are shown as orange dashed lines. Data for SN 2007ru are obtained from \citet{Sahu2009} and \citet{Wang2017}, and data for SN 2009bb are taken from \citet{Pignata2011}. }
\label{SN_LC}
\end{figure}

SN 2009bb is a Type Ic-BL SNe at $d_L \approx 40\mbox{ Mpc}$, first detected by the CHilean Automated Supernova sEarch (CHASE) on 2009 March 21. The optical and spectroscopic observations are carried out by \citet{Pignata2011}. The peak absolute magnitude of SN 2009bb is $M_B = -18.3\pm{0.44}\mbox{ mag}$, placing it in the bright end of the SNe Ic sequence. The Very Large Array (VLA) and the Chandra telescopes have also detected luminous radio emission and weak X-ray emission associated with SN 2009bb \citep{Soderberg2010}. Radio observations from SN 2009 suggest that the explosion required very high energy and a relativistic jet, which all point to an active central engine. 
%Simulations of relativistic jet-driven SNe have been carried out by a subset of authors \citep{Lazzati2012,Gilkis2014,Soker2017,Barnes2018,Shankar2021,Eisenberg2022,Wang2024}. %The simulation results show that jet will alter the velocity distribution of the SN ejecta \citep{Eisenberg2022}, but the contribution to the thermal evolution of the ejecta is negligible due to its short duration.  
In the magnetar model, the injected energy from the magnetar spin-down wind is dissipated either by channeling the jet or thermalizing the ejecta \citep{Wang2024}. We distribute the energy between the jet and the thermalized ejecta by introducing an efficiency factor $\xi$ in Eq \ref{eq:energy}.
%Therefore, jet would not significantly change the energy budget in Eq.\ref{eq:energy}, but affects the time scale of radiative dissipation. For SN 2009bb, it is not clear whether the jet will effectively accelerate the ejecta, thus we have not considered the contribution of the jet to the ejecta.}
We use a single power-law function to fit the light curve after the luminosity peak, and obtain the decay slope of $-1.65$, which suggests that the magnetar spin-down is caused by a combination of GW quadrupole and MD torque. 
We used the magnetar models with different spin modes to fit the luminosity curve of SN 2009bb, and the results are shown in Fig \ref{SN_LC}. The most suitable model for SN 2009bb is the hybrid magnetar model with the GW and MD acting together on the spin-down. The best-fitting results are shown in Table \ref{parameters}. In addition, we match the observed data of SN 2009bb with the $^{56}$Ni cascade decay model and find that the $^{56}$Ni model can explain its early light curve, but it does not agree with the later data, since the decay rate of the light curve is slower than expected by the $^{56}$Ni model.

Modeling the light curve of SN 2009bb, we derive parameters for magnetar and SN explosions, such as $B\sim 1.35\times 10^{15} \mbox{ G}$, $P_0\sim 26.86 \mbox{ ms}$, $\epsilon\sim 5.33\times 10^{-3} $, $M_{\rm ej}\sim 1.08 M_{\sun}$, $E_{\rm sn}\sim 2.91\times 10^{51}\mbox{ erg}$. The values of $B$ and $P_0$ are reasonable parameters for magnetar-driven SNe Ic-BL \citep{Kashiyama2016}. The corner plots in Fig 5 show that the ellipticity cannot be tightly constrained, so we give the upper and lower limits of $1\sigma$ confidence interval. The main reason for this may be that the energy lost by GW radiation is not prominent compared to the energy lost by MD radiation. The value of $E_{\rm sn}$ is comparable to the energy of a neutrino-driven SN explosion. 
Having such a good fitting to the light curve, we argue that SN 2009bb may be powered by a new-born magnetar with deformation.

\subsection{SN 2007ru}

SN 2007ru is a nearby (redshift $z = 0.01546$) SNe Ic-BL that was detected on 2007 December 2 by the Himalayan Faint Object Spectrograph Camera (HFOSC) \citep{Sahu2009}. It peaked at $8$ days after its discovery, and the peak magnitude is $M_V \approx -19.09 \mbox{ mag}$, making it one of the brightest SNe Ic. Its light curve shows no obvious sign of undulation, decaying with a slope of $-1.75$ after the peak luminosity, which is consistent with the scenario for both GW and MD acting together for the magnetar spin-down. 

By assuming that the SN 2007ru was powered by $^{56}$Ni cascade decay, \citet{Sahu2009} has inferred the $^{56}$Ni mass and ejecta mass of $M_{\rm Ni}\approx 0.4 M_{\sun}$ and $M_{\rm ej}\approx 1.3 M_{\sun}$, respectively. The ratio between $M_{\rm Ni}$ and $M_{\rm ej}$ is larger than the upper limit for CCSNe, which implies that the energy source is unlikely to be $^{56}$Ni decay. Fig \ref{SN_LC} shows the fitting results for different spin-down models and $^{56}$Ni model. The best-fitting model to SN 2007ru is the magnetar model with a combination of GW radiation and MD radiation. The other models cannot reproduce its early or late data. These results suggest that the bolometric luminosity evolution of SN 2007ru contains a contribution from GW radiation. \citet{Omand2024} used the optical emission from SN 2007ru to constrain the braking index of the NSs under the assumption of the magnetar model and achieved results consistent with ours.

Comparison of SN 2007ru observations with magnetar models allows us to constrain the parameters of magnetar and SN explosion as follows $B\sim 1.26\times 10^{15} \mbox{ G}$, $P_0\sim 25.19 \mbox{ ms}$, $\epsilon\sim 4.79\times 10^{-3} $, $M_{\rm ej}\sim 1.05 M_{\sun}$, $E_{\rm sn}\sim 27.81\times 10^{50}\mbox{ erg}$. The derived best-fit ellipticity has a large uncertainty.
The values of $B$ and $P_0$ are consistent with previous results for the magnetar models presented by \citep{Wang2017}, but there is a discrepancy in the mass of the ejecta. This difference is caused by the fact that they took a fixed value for the optical depth, i.e., $\kappa$=0.1 cm$^2$~g$^{-1}$ and then got the mass of ejecta of $M_{\rm ej}$, smaller than that of the SLSNe but consistent with stripped SNe. 
Our fitting results suggest that SN 2007ru is produced by a stripped SNe Ic-BL and probably powered by a rapidly spinning, deformed magnetar.

Note that so far no magnetar candidates for a GW-dominated spin-down have ever been found in SNe observations for two possible reasons. One is that GW-dominated spin-down may not play a significant role in the evolution of magnetars; the other is that the number of SNe observed is currently insufficient to find a better case. 
We find that the ellipticity of magnetar should be around $10^{-3}$ for modeling the light curves of SNe 2009bb and 2007ru. This value is consistent with the conclusions of \citet{Moriya2016} and \citet{Ho2016} on the upper limit of the ellipticity obtained by a simple comparison of the timescales of GW radiation and MD radiation for the magnetar-driven SNe. Previously magnetar-driven GRB models have also taken an ellipticity around $10^{-3}$ \citep{Lasky2016,xie2022}, which may suggest that the deformed magnetars are the sources for both GRBs and SNe Ic-BL.

\begin{figure}
 \centering
\includegraphics[width=0.49\textwidth]{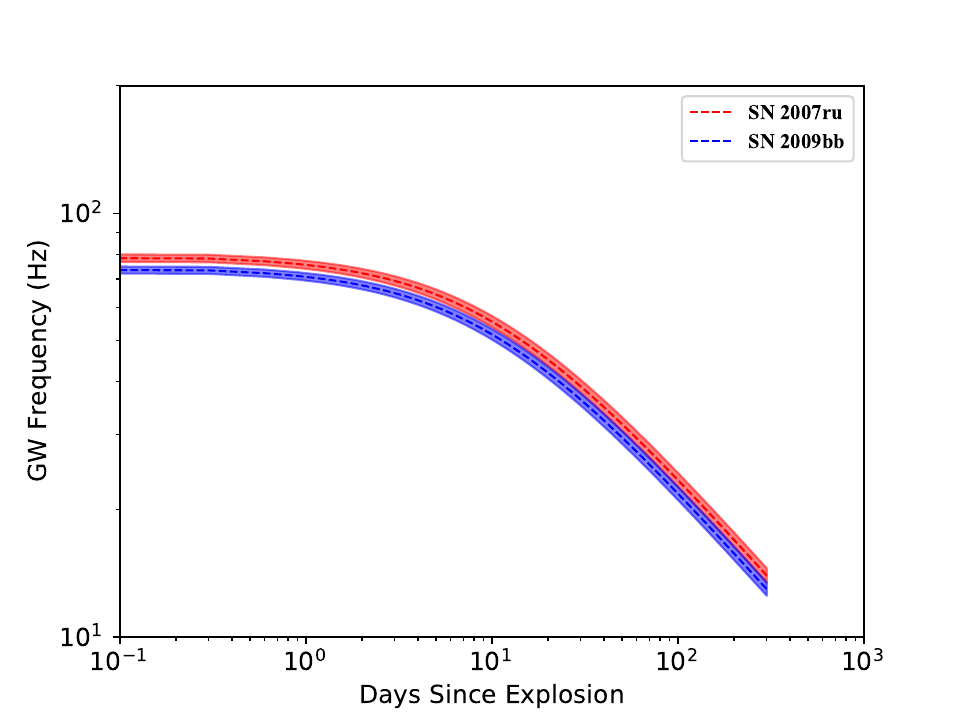}
\includegraphics[width=0.49\textwidth]{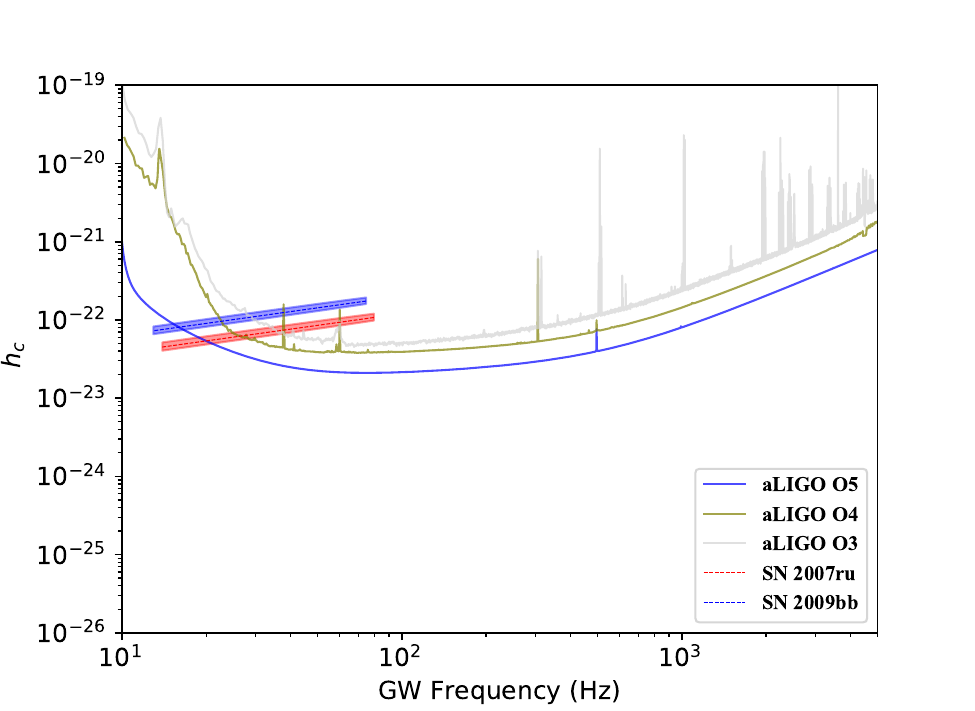}
\caption {Left panel: Frequency evolution of the predicted GW for the SNe 2007ru and 2009bb. The bands are the error range of the GW frequency. Right panel: The GW amplitude evolution for two SNe by assuming that the center engine is a magnetar. The bands with colors represent the error ranges. The designed sensitivity of O3, O4, and O5 of aLIGO are shown as light grey, olive, and blue curves, respectively. The projected sensitivity of the ET is shown as a black dashed curve. }
\label{hc}
\end{figure}

\section{Detectability of the GW signals from the deformed magnetars}

The current aLIGO and Virgo detectors have discovered more than 90 cases of compact star mergers with high confidence, which have opened a new era of GW astronomy. In addition to the merger of NSs and/or black holes, the continuous GWs from isolated deformed NSs are also important targets for aLIGO and Virgo detectors. Post-collapse magnetars are considered to be the most promising class of NSs for detecting continuous GWs due to their extremely strong magnetic fields and fast spinning. If the magnetars are the central engines of SNe, they would emit observable GWs due to non-axisymmetric deformation.

The GW strain from spinning magnetar with a mass quadrupole moment is \citep{Corsi2009, Howell2011, Dall'Osso2015}
\begin{eqnarray}
h(t)=\frac{4G\Omega(t)^2}{c^4d}I\epsilon,
\label{eq:ht}
\end{eqnarray}
Here $d$ is the distance from the source. For long-lived GW transient sources, the corresponding optimal matched filter signal-to-noise ratio is \citep{Corsi2009}
\begin{eqnarray}
	\nonumber \rho^2_{max}=\int^{+\infty}_{0} \left(\frac{h_c}{h_{rms}}\right)^{2} d(\ln f),
	\label{eq:rho}
\end{eqnarray}
where $h_c=f h(t)\sqrt{dt/df}$ is the GW amplitude, $h_{rms}=\sqrt{f S_h(f)}$ is the noise curve of detector, $S_h(f)$ is the power spectral density of the detector noise.
Using Eq \ref{eq:ht} to solve for $h_c$, one can obtain the GW amplitude of the magnetar as being \citep{Corsi2009}
\begin{equation}
h_c= \frac{1}{d}\sqrt{\frac{5GIf}{2c^3}},
\label{eq:hc}
\end{equation}
where $f=\Omega/\pi$ is the frequency of the GW signal.

According to Eq \ref{eq:hc} we can see that the $h_c$ is not only related to the evolution of the stellar spin frequency but also depends on the distance to a source. By fitting the SNe luminosity curve with the magnetar model, we can derive the evolution of the spin frequency and hence the $h_c$.  
The left side of Fig \ref{hc} shows the frequency evolution of the predicted GW for the SN 2007ru and SN 2009bb. From Eq \ref{eq:omegadot} and Eq \ref{eq:hc} it can be seen that the fractional uncertainty in the magnetar parameters will be directly converted to the uncertainty on the GW frequency and $h_c$.
The corner plots show that the ellipticity has a large uncertainty, which increases the uncertainties of the inferred GW frequency and $h_c$. We plot the $h_c$ curves of two SNe on the right side of Fig \ref{hc}, together with the sensitivity curves of aLIGO and Einstein detector (ET) \citep{abbott2018}. 
The uncertainties of the magnetar parameters have a small effect on the evolution of the $h_c$ curve. One can find that the GWs from SN 2009bb and SN 2007ru can reach the detection threshold of the aLIGO and ET detectors. On the other hand, for magnetars not yet detected by ground-based detectors for GWs, we can give an upper limit on the $\epsilon$ by comparing the sensitivity of GW detectors and $h_c$. Note that our estimate of the detectability of the GW signal is based on the assumption of an optimally matched filter, which may not be feasible for the detection of the proposed signals. The $h_c$ curves in Fig \ref{hc} represent the most optimistic upper limit for GW detection.

\section{Conclusion and Discussion}

A rapidly spinning magnetar can be formed after the SN explosion, emitting continuous GWs due to the magnetic-induced deformation or fluid oscillation excited by fast spinning. We explore the GW radiation in the magnetar-driven SNe model and try to find evidence of GW radiation in the SN light curves. Both MD radiation and GW radiation can act for the spin-down of magnetars. The GW energy loss can lead to a reduced MD radiative energy to heat the SN, causing a reduced peak luminosity of the SN and a fast slowing down, and also a low decay rate of light curve tails. We made magnetar models with GW emission and reproduced the light curves of SN 2009bb and SN 2007ru. We derived the parameters of deformed magnetars by fitting the observed light curves of SNe, and also estimated the detectability of the GWs from deformed magnetars. Our results suggest that the GWs from SN 2009bb and SN 2007ru can reach the detection threshold of the aLIGO and ET detectors. 
If GWs associated with SNe are detected in the future, this would be a smoking gun that magnetars can power SNe.

The effect of long-lived magnetar remnants on SN has been explored in detail \citep[e.g.][]{Inserra2013, Kasen2016}. The rotational energy reservoir of magnetars has been used to explain many different features of SN, such as the high peak luminosity, the large ejecta kinetic energy, and the tail decaying with $t^{-2}$. However, in some cases, SNe with such a tail may be attributed to the magnetar plus $^{56}$Ni model. The magnetar model explains the early fast rise and peak luminosity, while the $^{56}$Ni model explains the late tails, depending on if sufficiently massive Ni is synthesized. It may be difficult to synthesize more than $0.2 M_{\sun}$ of $^{56}$Ni  \citep{Nishimura2015, Suwa2015} in the SN with a magnetar. Our simulations suggest that a magnetar model plus $^{56}$Ni model does have some difficulty in explaining the overly flat decay of the tail (slope much less than $-2$), especially for decays longer than 300 days because the decay rate of $^{56}$Co gradually deviates over time.

By modeling the SN bolometric luminosity curve, we derive the ellipticity of the magnetar to be $\epsilon \sim 10^{-3}$. The current GW detection has set an upper limit of the ellipticity of $\sim 10^{-4}-10^{-6}$ for pulsar \citep{abbott2019a}, and $\sim 10^{-3}$ for young NSs \citep{Aasi2015b, Aasi2016}; while the electromagnetic observations from the GRB X-ray plateau constrain the ellipticity of magnetars to $\sim 10^{-3}-10^{-4}$ \citep{Lasky2016,xie2022}. Note, however, that new-born magnetars may have greater non-axisymmetric deformation than old pulsars, and thus may produce stronger GW emission.

The ellipticity of a magnetar inferred from two SNe are consistent with those inferred from the GRB observations, implying that both GRB and some SNe Ic-BL may have similar central engines. The deformation of NSs may be induced by anisotropic stresses caused by the internal magnetic field \citep{Cutler2002, Mastrano2011, Lander2014}. If NS distortion is induced by the pure toroidal magnetic field ($B_{\rm t}$), the relation between the $\epsilon$ and $B_{\rm t}$ can be described as $\epsilon\approx 1.6\times 10^{-4}({B_{\rm t}}/{10^{16}\,{\rm G}})^{2}$ \citep{Cutler2002, Stella2005, Dall'Osso2009}.
It can be seen that the $B_{t}$ ought to reach $\sim 2\times10^{16}\mbox{ G}$ for the $\epsilon$ of $\sim 10^{-3}$, which implies that a stronger toroidal field strength is required,  at least $1-2$ orders of magnitude larger than the dipole field ($B\sim 10^{14}-10^{15} \mbox{ G}$). Theoretically, the toroidal magnetic field of magnetar can be amplified to $\sim 10^{16}-10^{17}\mbox{ G}$ due to the $\alpha-\Omega$ dynamo mechanism \citep{Duncan1992}. This conclusion is supported by previous works on constraining the internal field strength of soft gamma-ray repeating bursts (see, \citep{Ioka2001, Corsi2011}).  Other channels than magnetic-induced NS deformation, such as secular bar-mode instability (also known as unstable f-mode oscillation), can also play an important role in nascent magnetars. \citet{Lasky2016} proposed that the value of the maximum ellipticity induced by f-mode depends on the saturation energy of the mode. The maximum energy of the f-mode is $ E_{\rm mode} \sim 10^{-6} M c^{2}$, which corresponds to the effective ellipticity of $ \sim 2 \times 10^{-3}$. In addition, NS accretion can also induce distortion. The material around the accretion disc will flow into the poles of the NS due to the gravitational effect to form mountains. In this scenario, the magnitude of the ellipticity value of the NS depends on the mass of the accreting material \citep{Zhong2019}. \citet{Sur2021} suggested that for accreting NSs located at $1 \mbox{ Mpc}$, the GW amplitudes at kHz frequencies can reach the detection threshold of aLIGO. 

Even though aLIGO has not detected continuous GWs from isolated NSs, we are optimistic about the prospect of detecting GWs from SNe. If continuous GWs from the NSw can be detected by future GW detectors, then we can constrain the ellipticity $\epsilon$ of the NS by Eq \ref{eq:ht}. Furthermore, the duration and frequency of continuous GWs can be obtained from their waveforms. According to Eq \ref{eq:Lem}, \ref{eq:Edotgwe} and \ref{eq:hc}, the mass-radius relation can be constrained and then combined with the inferred ellipticity $\epsilon$ to provide valuable information for constraining the NS deformation after the SN explosion.

\begin{figure}
 \centering
\includegraphics[width=0.99\textwidth]{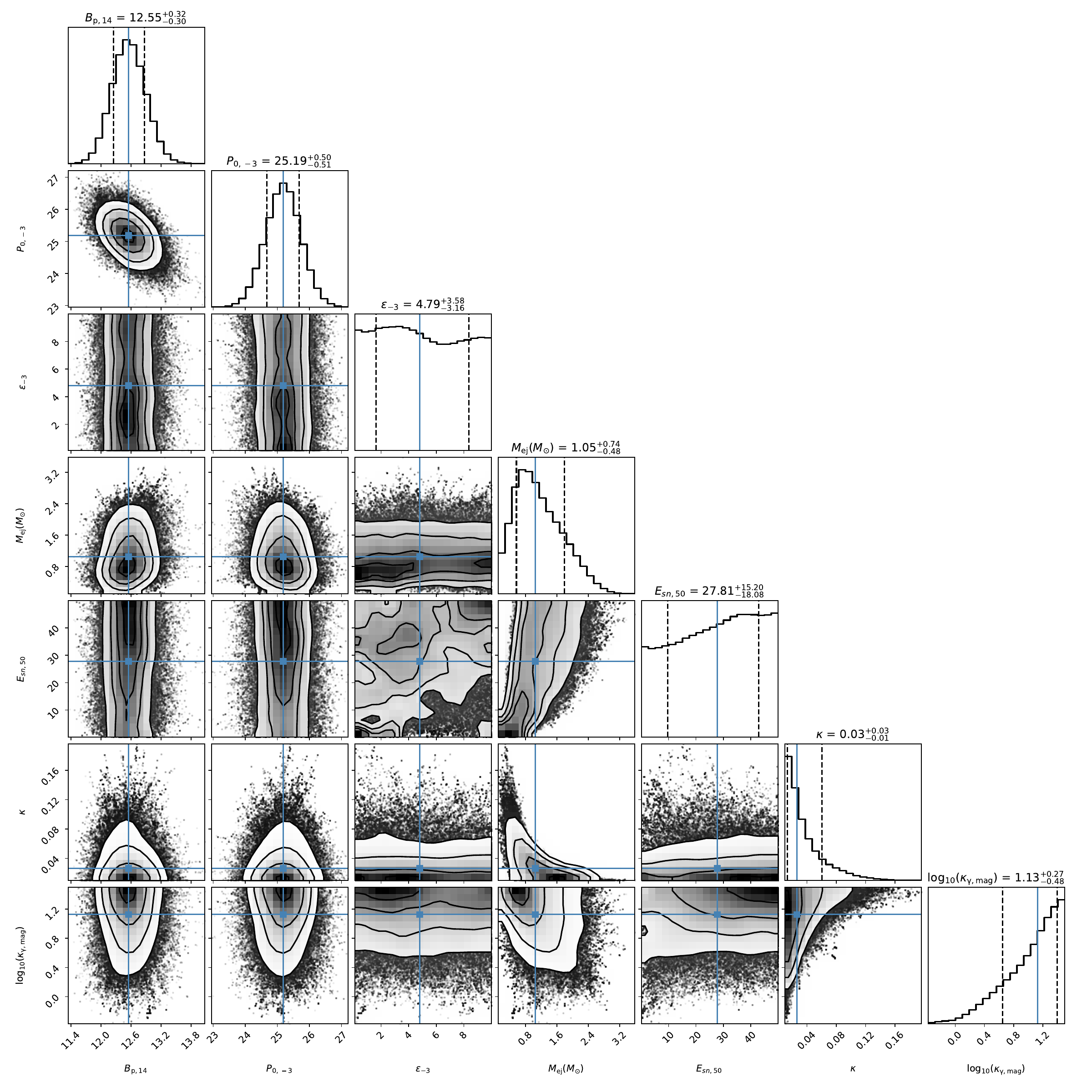}
\caption {The corner plot of the magnetar model for supernova 2007ru. The $1\sigma$ confidence level of the parameters is shown as the vertical dashed lines. }
\label{SN_LC2007RU}
\end{figure}

\begin{figure}
 \centering
\includegraphics[width=0.99\textwidth]{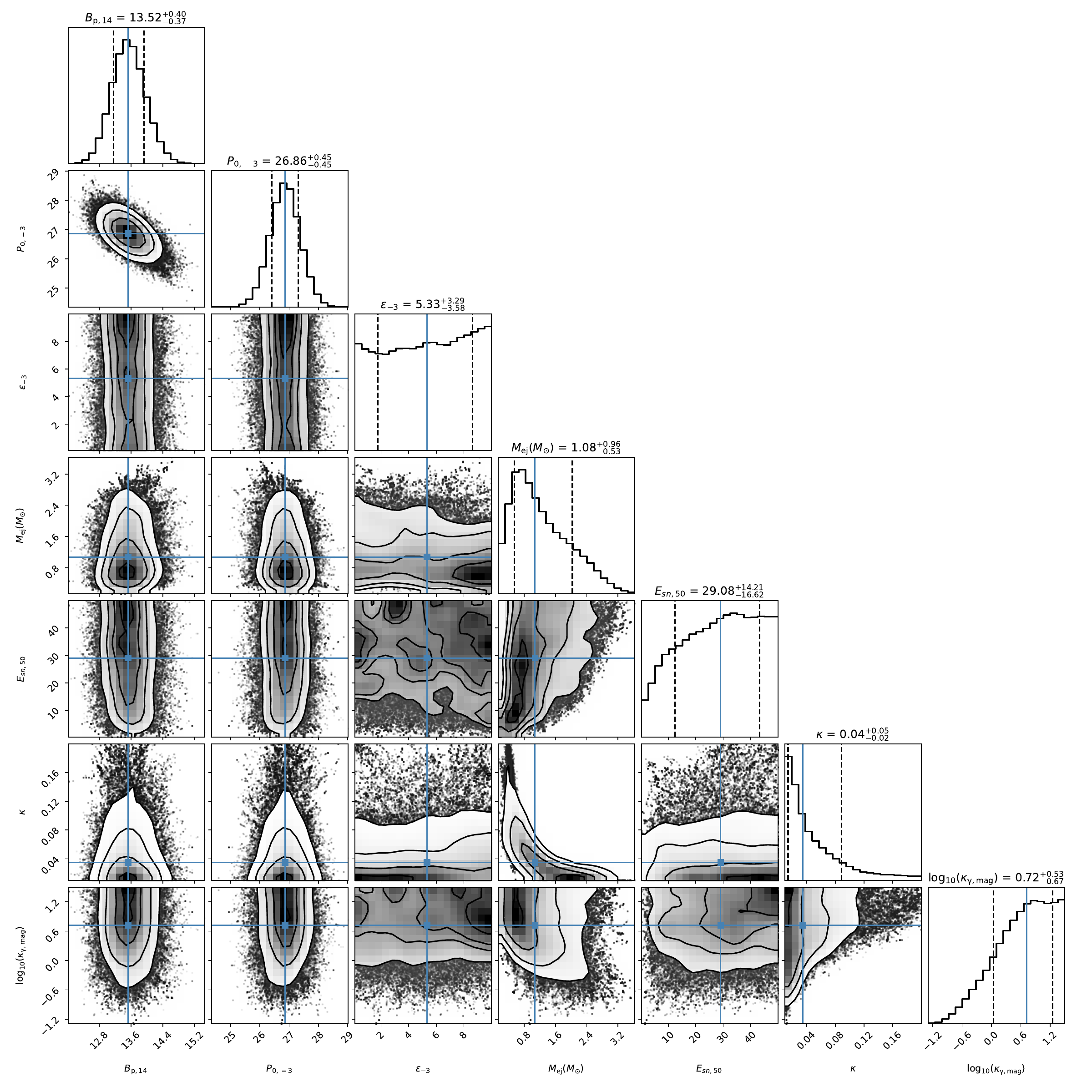}
\caption {The corner plot of the magnetar model for supernova 2009bb. }
\label{SN_LC2009BB}
\end{figure}

\section*{ Acknowledgements }

We thank the anonymous referee for helpful comments to improve this paper.
The authors are supported by the National Natural Science Foundation of China (Nos. 11988101, 11833009, 12073080, 12233011, 11933010, 11921003 and 12303050) and by the Chinese Academy of Sciences via the Key Research Program of Frontier Sciences (No. QYZDJ-SSW-SYS024), China Postdoctoral Science Foundation (grant No. 2023M743397), and the Fundamental Research Funds for the Central Universities.

\bibliography{V1}
\bibliographystyle{aasjournal}
\end{document}